\documentclass{amsart}

\usepackage{t1enc}
\usepackage[utf8]{inputenc}

\usepackage{latexsym,pifont}
\usepackage{epsfig}
\usepackage{rotating}

\usepackage{ifpdf}
\ifpdf
\usepackage{epstopdf}
\usepackage{hyperref}
\else
\usepackage[hypertex]{hyperref}
\fi

\usepackage{amsfonts,amsma th,amssymb,mathrsfs,extarrows,MnSymbol}
\usepackage[all]{xy}

\usepackage{tikz}
\usetikzlibrary{matrix,arrows}

\newcommand{\brem}{\begin{Rem}}
\newcommand{\erem}{\end{Rem}\medskip}
\newcommand{\beg}{\begin{Eg}}
\newcommand{\eeg}{\end{Eg}}
\newcommand{\bedef}{\begin{Def}}
\newcommand{\exdef}{\begin{flushright}$\diamond$\end{flushright}
\end{Def}\vskip0.1cm}
\newcommand{\berop}{\begin{Prop}}
\newcommand{\eerop}{\end{Prop}}
\newcommand{\belem}{\begin{Lem}}
\newcommand{\elem}{\end{Lem}}
\newcommand{\bethe}{\begin{Thm}}
\newcommand{\ethe}{\end{Thm}}
\newcommand{\becor}{\begin{Cor}}
\newcommand{\ecor}{\end{Cor}}
\newcommand{\beroof}{\noindent\begin{proof}}
\newcommand{\eroof}{\end{proof}}
\newcommand{\becon}{\begin{Conv}}
\newcommand{\econ}{\begin{flushright}$\checkmark$\end{flushright}\end{Conv}}
\newcommand{\befact}{\begin{Fact}}
\newcommand{\efact}{\begin{flushright}$\checkmark$\end{flushright}\end{Fact}}
\newcommand{\bequest}{\begin{Quest}}
\newcommand{\equest}{\end{Quest}}
\newcommand{\brob}{\begin{Prob}}
\newcommand{\erob}{\end{Prob}}
\newcommand{\becj}{\begin{conj}}
\newcommand{\ecj}{\begin{flushright}$\boxtimes$\end{flushright}\end{conj}}

\newcommand{\barr}{\begin{array}}
\newcommand{\earr}{\end{array}}
\newcommand{\ben}{\begin{enumerate}}
\newcommand{\een}{\end{enumerate}}
\newcommand{\bit}{\begin{itemize}}
\newcommand{\eit}{\end{itemize}}

\newcommand{\qq}{\begin{eqnarray}}
\newcommand{\qqq}{\end{eqnarray}}

\newcommand{\nn}{\nonumber}

\newcommand{\ovl}[1]{\overline{#1}}
\newcommand{\unl}[1]{\underline{#1}}

\newcommand\void[1]{}

\newcommand{\gt}[1]{\mathfrak{#1}}

\def\cA{\mathcal{A}}
\def\cB{\mathcal{B}}
\def\cC{\mathcal{C}}

\def\cF{\mathcal{F}}
\def\cG{\mathcal{G}}

\def\cI{\mathcal{I}}
\def\cJ{\mathcal{J}}
\def\cK{\mathcal{K}}

\def\cM{\mathcal{M}}

\def\cO{\mathcal{O}}

\def\cS{\mathcal{S}}
\def\cT{\mathcal{T}}

\def\cV{\mathcal{V}}

\def\xcC{\mathscr{C}}

\def\bC{{\mathbb{C}}}

\def\bN{{\mathbb{N}}}

\def\bR{{\mathbb{R}}}
\def\bS{{\mathbb{S}}}

\def\bZ{{\mathbb{Z}}}

\def\a{\alpha}

\def\G{\Gamma}

\def\D{\Delta}

\def\vep{\varepsilon}

\def\k{\kappa}

\def\la{\lambda}
\def\La{\Lambda}
\def\om{\omega}
\def\Om{\Omega}

\def\si{\sigma}
\def\Si{\Sigma}

\def\Bgt{\gt{B}}

\def\ggt{\gt{g}}

\newcommand{\sfi}{{\mathsf i}}

\newcommand{\sfk}{{\mathsf k}}

\newcommand{\sfN}{{\mathsf N}}

\newcommand{\sfP}{{\mathsf P}}

\newcommand{\txD}{{\rm D}}
\newcommand{\ee}{{\rm e}}

\newcommand{\txg}{{\rm g}}
\newcommand{\txG}{{\rm G}}

\newcommand{\txm}{{\rm m}}

\newcommand{\txN}{{\rm N}}

\def\exp{{\rm exp}}
\def\id{{\rm id}}
\newcommand{\pr}{{\rm pr}}

\def\too{\longrightarrow}

\def\Hom{{\rm Hom}}

\def\1morf{1{\rm -Mor}}
\def\2morf{2{\rm -Mor}}
\def\dim{{\rm dim}}

\def\bgrb{\gt{BGrb}}

\newcommand{\sMan}{{\rm {\bf sMan}}}

\def\Vol{{\rm Vol}}

\def\p{\partial}

\def\emb{\hookrightarrow}

\def\Hol{{\rm Hol}}

\def\bd1{{\boldsymbol{1}}}
\def\brd0{{\boldsymbol{0}}}

\def\tr{{\rm tr}}

\def\Ad{{\rm Ad}}

\def\x{\times}
\def\ox{\otimes}

\def\rx{\rtimes}

\def\rstr{\mathord{\restriction}}

\newtheorem{Thm}{Theorem}
\newtheorem{Prop}[Thm]{Proposition}
\newtheorem{Lem}[Thm]{Lemma}
\newtheorem{conj}{Conjecture}
\newtheorem{Cor}[Thm]{Corollary}
\theoremstyle{definition}
\newtheorem{Rem}[Thm]{Remark}
\newtheorem{Def}[Thm]{Definition}
\newtheorem{Eg}[Thm]{Example}
\newtheorem{Conv}[Thm]{Convention}
\newtheorem{Fact}[Thm]{Fact}
\newtheorem{Quest}[Thm]{Question}
\newtheorem{Prob}[Thm]{Problem}

\begin{document}
\title{Towards higher super-$\si$-model categories}

\author{Rafa\l ~R.\ ~Suszek}
\address{R.R.S.:\ Katedra Metod Matematycznych Fizyki,\ Wydzia\l ~Fizyki
Uniwersytetu Warszawskiego,\ ul.\ Pasteura 5,\ PL-02-093 Warszawa,
Poland} \email{suszek@fuw.edu.pl}

\begin{abstract}
A simplicial framework for the gerbe-theoretic modelling of su\-per\-charged-loop dynamics in the presence of  worldsheet defects is discussed whose equivariantisation with respect to global supersymmetries of the bulk theory and subsequent orbit decomposition lead to a natural stratification of and cohomological superselection rules for target-space supergeometry,\ expected to encode essential information on the quantised 2$d$ field theory.\ A physically relevant example is analysed in considerable detail.
\end{abstract}

\maketitle

\section{Introduction}\label{sec:intro}

Models of geometrodynamics of extended distributions of (super)charge,\ known as nonlinear $\si$-models with topological Wess--Zumino terms,\ remain an active field of physical and mathematical research,\ not least because of their applicability in the description of a wide range of dynamical systems,\ from the critical field theory of collective excitations of quantum spin chains all the way to classical superstring theory,\ and because of the great wealth of the underlying mathematics.\ The hierarchical higher-geometric structures -- a.k.a.\ $p$-gerbes and their morphisms -- associated with background $(p+2)$-form fields in these models,\ which couple to the worldvolume (super)charge current through distinguished Cheeger--Simons differential characters generalising the standard line holonomy (as derived for $\,p=1\,$ in \cite{Gawedzki:1987ak,Gawedzki:2002se,Fuchs:2007fw,Runkel:2008gr}),\ have long been known not only to give rise to a canonical prequantisation of the models,\ as in \cite{Gawedzki:1987ak,Suszek:2011hg},\ and to provide a natural cohomological classification of the models themselves,\ their boundary conditions and generic defects,\ but also -- to lead to a categorification of their prequantisable group-theoretic symmetries,\ including the gauged ones,\ and more general dualities ({\it e.g.},\ $T$-duality).\ These correspondences can oftentimes be implemented by certain defect networks embedded in the $\si$-model spacetime,\ to which sheaf-theoretic data of $k>0$-cells of the weak $(p+1)$-categories with $p$-gerbes as 0-cells are pulled back along the $\si$-model field in order to render the Dirac--Feynman amplitude of the $\si$-model with defects well-defined,\ {\it cp.}\ \cite{Runkel:2008gr,Runkel:2009sp,Suszek:2011hg,Suszek:2012ddg}.\ In particular,\ the distinguished \emph{topological} defects,\ studied at some length in \cite{Frohlich:2006ch,Fuchs:2007fw,Runkel:2008gr,Runkel:2009sp,Suszek:2011hg},\ enter the generalised world-sheet orbifold construction of \cite{Frohlich:2006ch,Runkel:2008gr} of the $\si$-model on the space $\,M/\txG_\si\,$ of orbits of an action of a rigid-symmetry group $\,\txG_\si\,$ on the target space $\,M\,$ of the original $\si$-model,\ in which they serve to model the so-called $\txG_\si$-twisted sector of the orbifold theory,\ instrumental in the gauging of $\,\txG_\si$.\ Finally,\ in some highly symmetric settings,\ such as,\ {\it e.g.},\ that of the Wess--Zumino--Witten (WZW) $\si$-model,\ there are defects realising the \emph{generating} symmetries of the $\si$-model \cite{Fuchs:2007fw,Runkel:2008gr,Runkel:2009sp,Suszek:2022lpf} which encode highly nontrivial information on the ensuing quantum field theory,\ such as,\ {\it e.g.},\ its fusion rules and Moore--Seiberg data.\ Therefore,\ it is clear that understanding the structure of the fully fledged higher categories behind the topological couplings of poly-phase $\si$-models,\ and constructing concrete examples of their $k$-cells,\ is a goal of fundamental relevance to the study of these important field theories.

This note gives a concise account of a proposal,\ advanced in \cite{Suszek:2022lpf} on the basis of the earlier studies \cite{Runkel:2008gr,Runkel:2009sp,Suszek:2011hg,Suszek:2012ddg},\ for an effective structurisation of the stratified target spaces and the higher-geometric objects over them defining the 2$d$ $\si$-model in the presence of defects compatible with configurational symmetries of the bulk field theory.\ The proposal is formulated in the most general (target-space) $\bZ/2\bZ$-graded setting of the Green--Schwarz-type super-$\si$-model and uses mixed group-theoretic,\ simplicial and cohomological tools.\ It paves the way to a systematic construction of maximally supersymmetric defects in the flat $\bZ/2\bZ$-graded target geometry,\ and leads to interesting novel predictions for a higher-geometric target-space realisation of the non-perturbative data of the bulk theory in the highly (super)symmetric setting of the WZW(-type) models with Lie-(super)group targets.

\section{A bicategory for the super-$\si$-model with defects}\label{sec:bicat}

The 2$d$ super-$\si$-model is a lagrangean theory of (super)fields from the mapping supermanifold $\,[\Si,M]=\unl{\rm Hom}_\sMan(\Si,M)\,$ defined for a closed oriented (Gra\ss mann-)even manifold $\,\Si\,$ (the `spacetime' of the model) and a supermanifold $\,M$.\ For $\,M\,$ even,\ $\,[\Si,M]\,$ is represented by $\,C^\infty(\Si,M)\,$ and we obtain a theory of embeddings shaped by an interplay between forces sourced by two tensor fields on the target space $\,M$:\ a metric tensor $\,\txg\,$ and a closed 3-form Kalb--Ramond field $\,\chi\,$ with periods from $\,2\pi\bZ$.\ For $\,M\,$ properly graded,\ the inner-$\Hom$ functor $\,[\Si,M]\,$ is to be evaluated on the nested family of superpoints $\,\bR^{0|N},\ N\in\bN^\x$,\ both tensors are even,\ and the `metric' $\,\txg\,$ typically degenerates in the Gra\ss mann-odd directions.\ In either setting,\ the 3-form geometrises,\ in the sense of \cite{Murray:1994db},\ as a gerbe $\,\cG\,$ with connective structure of curvature $\,\chi$.\ The gerbe trivialises upon pullback to the 2$d$ worldsheet $\,\Si$,\ thereby defining -- for $\,\p\Si=\emptyset\,$ -- the topological Wess--Zumino (WZ) coupling of $\,\chi\,$ to the charged-loop current in $\,M$,\ given by the surface holonomy \cite{Gawedzki:1987ak} of $\,\cG\,$ along the image of $\,\Si\,$ in $\,M$.

In the boundary $\si$-model,\ with $\,\p\Si\neq\emptyset$,\ specifying the gerbe $\,\cG\,$ alone is \emph{not} sufficient to obtain a consistent field theory as the topological term becomes ill-defined.\ Instead,\ a distinguished 1-cell from the bicategory $\,\bgrb_\nabla(M)\,$ of (1-)gerbes over $\,M\,$ is required to exist over a submanifold $\,\iota_D:D\emb M\,$ into which $\,\p\Si\,$ is constrained to map,\ to wit,\ a trivialisation $\,\cT_D:\iota_D^*\cG\cong\cI_\om\,$ in terms of a trivial gerbe $\,\cI_\om\,$ associated with a \emph{global} de Rham primitive $\,\om\in\Om^2(D)\,$ of $\,\chi$,\ {\it cp.}\ \cite{Gawedzki:2002se}.\ The trivialisation is to be pulled back to $\,\p\Si\,$ by the $\si$-model field. 

The boundary $\si$-model can be viewed as a two-phase field theory in which the bulk phase,\ defined by the triple $\,(M,\txg,\cG)\,$ abuts onto the empty phase $\,(\bR^{0|0},0,\cI_0)\,$ across the boundary domain wall (or boundary \emph{defect}) $\,\p\Si$.\ In this picture,\ the target space is the \emph{stratified} supermanifold $\,\widetilde M\equiv M\sqcup\bR^{0|0}$,\ and the two limiting field configurations at the defect are modelled by the two mappings $\,\iota_1\equiv\iota_D:D\too M\subset\widetilde M\,$ and $\,\iota_2:D\dashrightarrow\bR^{0|0}\subset\widetilde M$.\ This is the point of departure for a far-reaching generalisation,\ contemplated in \cite{Fuchs:2007fw,Runkel:2008gr},\ in which an arbitrary number of phases coexist over 2$d$ patches within $\,\Si$,\ separated by \emph{defect lines} which,\ in turn,\ intersect at a {\it discretuum} of \emph{defect junctions},\ graded by their valence.\ The physical prototype of this situation is the decomposition of a demagnetised ferromagnetic medium into Weiss domains of uniform magnetisation which jumps across domain walls between them,\ and the most natural application of defects in the field-theoretic setting under consideration is the modelling of the \emph{twisted sector} in the theory with the target space given by the orbispace of an action (not necessarily free or proper) of a symmetry group $\,\txG_\si\subset{\rm Isom}(M,\txg)\,$ of the $\si$-model for $\,[\Si,M]\,$ -- this orbifold $\si$-model can be \emph{defined} in terms of ($\txG_\si$-classes of) patchwise continuous field configurations in $\,M\,$ with $\txG_\si$-jump discontinuities localised at  an arbitrarily fine mesh of defect lines carrying data of a $\txG_\si$-equivariant structure on $\,\cG\,$ \cite{Jureit:2006yf,Runkel:2008gr,Frohlich:2009gb,Suszek:2012ddg}.

The above considerations set the stage for a precise definition of the poly-phase super-$\si$-model:\ It starts with a splitting of $\,\Si\in\p^{-1}\emptyset\,$ by an embedded defect graph $\,\G\subset\Si\,$ into a family $\,\{D_i\}_{i\in I}\,$ of (topologically) closed 2$d$ domains $\,D_i\,$ which compose the extended worldsheet $\,\widehat\Si=\sqcup_{i\in I}\,D_i\,$ and intersect at a family $\,\{L_{(i,j)}\}_{(i,j)\in I_\G\subset I^{\x 2}}\,$ of closed oriented defect lines $\,L_{(i,j)}$.\ The latter form the extended defect graph $\,\widehat\G=\sqcup_{(i,j)\in I_\G}\,L_{i,j}\subset\widehat\Si\x_\Si\widehat\Si\equiv\widehat\Si{}^{[2]}\,$ and join transversally at defect junctions whose ensemble $\,V=\sqcup_{\nu\geq 3}\,V_\nu\,$ decomposes into subsets $\,V_\nu\subset\widehat\Si{}^{[\nu]}\,$ of junctions of a fixed valence $\,\nu$,\ further split as $\,V_\nu=\sqcup_{\vep_\nu\in(\bZ/2\bZ)^{\x\nu}}\,V_{\vep_\nu},\ \vep_\nu=(\vep^{(\nu)}_{k,k+1}),$ $k\in\ovl{1,n}\,$ into subcomponents $\,V_{\vep_\nu}\,$ with distinct cyclic ($(\nu,\nu+1)\equiv(\nu,1)$) sequences of \emph{in}-coming ($\vep^{(\nu)}_{k,k+1}=+1$) and \emph{out}-going ($\vep^{(\nu)}_{k,k+1}=-1$) defect lines,\ {\it cp.}\ \cite{Runkel:2008gr}.\ There are canonical projections $\,p_1,p_2:\widehat\G\too\widehat\Si\,$ (assigning to a given defect line the corresponding components of the boundaries of the two domains separated by it) and $\,p^{\vep_\nu}_{k,k+1}:V_{\vep_\nu}\too\widehat\G\,$ (assigning to a given defect junction the endpoints of the defect lines converging at it,\ in anti-clockwise cyclic order),\ satisfying the obvious identities of order 2,\ {\it e.g.},\ $\,p_2\circ p^{(+++)}_{1,2}=p_1\circ p^{(+++)}_{2,3}\,$ for two defect lines $\,L_{1,2}\,$ and $\,L_{2,3}\,$ \emph{in}-coming at $\,\upsilon\in V_{(+++)}$.\ To the manifolds (with boundary) $\,\widehat\Si,\widehat\G\,$ and $\,V_\nu$,\ we associate the respective strata $\,M_0,M_1\,$ and $\,M_{\nu-1}=\sqcup_{\vep_\nu\in(\bZ/2\bZ)^{\x\nu}}\,T_{\vep_\nu}\,$ of the target supermanifold (possibly further stratified,\ as in the boundary example above),\ which determine a natural decomposition $\,\cF_\si\equiv[\widehat\Si,M_0]\sqcup[\widehat\G,M_1]\sqcup\sqcup_{\nu\geq 3}\,\sqcup_{\vep_\nu\in(\bZ/2\bZ)^{\x\nu}}\,[V_{\vep_\nu},T_{\vep_\nu}]\,$ of the space of $\si$-model fields.\ These target strata are endowed with smooth structure maps $\,\iota_1,\iota_2:M_1\too M_0\,$ and $\,\pi^{\vep_\nu}_{k,k+1}:T_{\vep_\nu}\too M_1\,$ subject to relations of order 2 mirroring those satisfied by their worldsheet counterparts,\ requisite for the consistency of the field-theoretic framework.\ The structure maps give rise to a family of pullback operators:\ $\,\D=\iota_2^*-\iota_1^*\,$ and $\,\D_{\vep_\nu}=\sum_{k=1}^\nu\,\vep^{(\nu)}_{k,k+1}\,\pi^{\vep_v\,*}_{k,k+1}\,$ which obey the identities $\,\D_{\vep_\nu}\circ\D=0\,$ and thus establish a relative-cohomological structure on $\,M\equiv M_0\sqcup M_1\sqcup\sqcup_{\nu\geq 3}\,\sqcup_{\vep_\nu\in(\bZ/2\bZ)^{\x\nu}}\,T_{\vep_\nu}\,$ in which the form $\,\Om\equiv(\chi,\om,0)\in\Om^3(M_0)\oplus\Om^2(M_1)\oplus\Om^1(M_{\nu\geq 3})\,$ acquires the status of a relative de Rham 3-cocycle,\ {\it cp.}\ \cite[Sec.\,7.2]{Suszek:2012ddg},\ and the defect strata $\,M_1\,$ and $\,T_{\vep_\nu}\,$ play the r\^ole of correspondence spaces:\ The former supports a trivialisation of the $\D$-image of the `bulk' gerbe $\,\cG\,$ on $\,M_0$,\ whereas the latter carries a secondary trivialisation of the $\D_{\vep_\nu}$-image of that primary trivialisation.\ In order to motivate this result of the in-depth analysis reported in \cite{Runkel:2008gr},\ let us consider the special case in which $\,\iota_{D^\x}:D^\x\equiv(\iota_1,\iota_2)(M_1)\emb M_0^{\x 2}\,$ is an embedding,\ and the connected component $\,\bS^1\subset\G\,$ mapped to $\,M_1\,$ separates diffeomorphic domains $\,D_1\,$ and $\,D_2$.\ Invoking the Wong--Affleck `folding trick' \cite{Wong:1994pa},\ we may then regard the defect line as a boundary defect in the $\si$-model on $\,D_1\,$ (onto which $\,D_2\,$ has been `folded') with the target supermanifold $\,M_0^{\x 2}\,$ and the gerbe $\,\pr_1^*\cG\ox\pr_2^*\cG^*\,$ (the dualisation of $\,\pr_2^*\cG\,$ reflects the flip of the orientation accompanying the `folding',\ {\it cp.}\ \cite{Gawedzki:2008um}),\ and with the boundary $\,\bS^1\,$ sent to $\,D^\x$.\ The reasoning of \cite{Gawedzki:2002se},\ referred to previously,\ now calls for a trivialisation $\,\pr_1^*\cG\ox\pr_2^*\cG^*\cong\cI_\om$,\ or,\ equivalently,\ a so-called gerbe bi-module $\,\Phi:\iota_1^*\cG\cong\iota_2^*\cG\ox\cI_\om\,$ over $\,M_1$.\ This turns out to be the right structure for an \emph{arbitrary} choice of the correspondence space $\,(M_1,\iota_\cdot)$,\ {\it cp.}\ \cite{Runkel:2008gr}.\ Note in the passing that there is always the distinguished \emph{identity defect},\ mapping to the component $\,M_0\subset M_1\,$ with $\,\iota_A\rstr_{M_0}\equiv\id_{M_0}\,$ and $\,\Phi\rstr_{M_0}\equiv\id_\cG$.\ There is no equally straightforward and general argument elucidating the gerbe-theoretic structure to be pulled back to the $\,V_{\vep_\nu}$,\ but we may give a heuristic reasoning which emphasises the main idea behind the rigorous construction while circumnavigating its technicalities,\ given {\it ibidem}.\ In it,\ we assume all lines to be \emph{in}-coming at $\,v\,$ for the sake of simplicity and drop the sign labels.\ Thus,\ whenever $\,I_{(\nu)}\equiv(\pi_{1,2},\pi_{2,3},\ldots,\pi_{\nu-1,\nu},\pi_{\nu,1})(T_{(++\cdots+)})\subset M_1^{\x\nu}\,$ is a submanifold,\ \emph{and} the value of the action functional of the $\si$-model is invariant under homotopy moves of the defect within $\,\Si$,\ {\it i.e.},\ when we are dealing with a \emph{topological} defect in a conformally invariant $\si$-model,\ we may -- upon cutting out a disc centred on a given $\,\upsilon\in V_{(++\cdots+)}\,$ whose boundary intersects each line emanating from $\,\upsilon\,$ once,\ and subsequently cutting out from it another disc of radius $\,r\approx 0\,$ with the same properties -- deform the defect lines $\,L_{k,k+1}\,$ on the ensuing annulus in such a way that the angular distance between the nearest neighbours $\,(L_{k,k+1},L_{k+1,k+2})\,$ is $\,\vep\approx 0(\approx\nu\vep)$.\ Now we should be able to read off the sought-after vertex structure from the limit $\,\vep\searrow 0\,$ succeeded by $\,r\searrow 0$.\ The former leaves us with a single defect line mapped into $\,M_1^{(\nu)}=\{(q_1,q_2,\ldots,q_\nu)\in M_1^{\x\nu}\ |\ \iota_2(q_k)=\iota_1(q_{k+1}),\ k\in\ovl{1,\nu-1}\}\,$ and carrying the data of the composite 1-isomorphism $\,(\Phi_\nu\ox\id_{\cI_{\om_1+\om_2+\cdots+\om_{\nu-1}}})\circ\cdots\circ(\Phi_3\ox\id_{\cI_{\om_1+\om_2}})\circ(\Phi_2\ox\id_{\cI_{\om_1}})\circ\Phi_1\,$ with $\,(\Phi_k,\om_k)=\pr^*_k(\Phi,\om)$,\ representing the \emph{fusion} of the $\,\nu\,$ gerbe bi-modules.\ The latter reproduces the endpoint $\,\widetilde\upsilon\,$ of the fused defect,\ or,\ equivalently,\ a junction between it and the identity defect stretching from $\,\widetilde\upsilon\,$ to the boundary of the big disc (and mapping to $\,M_0\subset M_1^{(\nu)}\,$ in a natural manner).\ In order to understand what ought to be put at the junction,\ we fold the disc along the fused defect \emph{and} its identity extension,\ whereupon we obtain a half-disc worldsheet with a \emph{piecewise} boundary condition at the fold (the other part of its boundary has a different status,\ and so we do not consider it here).\ At this stage,\ we may apply a dimensionally reduced variant of the folding trick to the boundary (Chan--Paton) degrees of freedom at the defect junction $\,\widetilde\upsilon\,$ mapped to $\,T_{(++\cdots+)}$,\ whereby it transpires that $\,\widetilde\upsilon\,$ should carry a trivialisation of $\,(\Phi_{\nu,1}\ox\id_{\om_{1,2}+\om_{2,3}+\cdots+\om_{\nu-1,\nu}})\circ\cdots\circ(\Phi_{3,4}\ox\id_{\cI_{\om_{1,2}+\om_{2,3}}})\circ(\Phi_{2,3}\ox\id_{\cI_{\om_{1,2}}})\circ\Phi_{1,2}\,$ for $\,(\Phi_{k,k+1},\om_{k,k+1})=\pi^*_{k,k+1}(\Phi,\om)\,$ (this makes sense as $\,\Phi\,$ is represented by a principal $\bC^\x$-bundle).\ The heuristic argument does not fix the (global) connection of the trivialisation,\ it is only the detailed computation of \cite{Runkel:2008gr} which shows that it should be null.\ Altogether,\ we end up with the superstring background $\,\Bgt=(\cM,\cB,\cJ)\,$ composed of:\ the bulk target $\,\cM=(M_0,\txg,\cG)\,$ in which $\,(M_0,\txg)\,$ is a quasi-metric supermanifold with a gerbe $\,\cG\,$ of curvature $\,\chi\,$ over it;\ the $\cG$-\emph{bi-brane} $\,\cB=(M_1,\iota_\cdot,\om,\Phi)\,$ with the bimodule $\,\Phi:\iota_1^*\cG\cong\iota_2^*\cG\ox\cI_\om$;\ and the $(\cG,\cB)$-\emph{inter-bi-brane} $\,\cJ=\sqcup_{\nu\geq 3}\,\sqcup_{\vep_\nu\in(\bZ/2\bZ)^{\x\nu}}\,(T_{\vep_\nu},\pi^{\vep_\nu}_{\cdot,\cdot},\varphi_{\vep_\nu})\,$ with the component fusion 2-isomorphisms
{\small\qq\nn
\varphi_{\vep_\nu}\ :\ \bigl(\Phi^{(\nu)}_{\nu,1}\ox\id_{\cI_{\D_{\vep_\nu}\om-\om^{(\nu)}_{\nu,1}}}\bigr)\circ\cdots\circ\bigl(\Phi^{(\nu)}_{3,4}\ox\id_{\cI_{\om^{(\nu)}_{1,2}+\om^{(\nu)}_{2,3}}}\bigr)\circ\bigl(\Phi^{(\nu)}_{2,3}\ox\id_{\cI_{\om^{(\nu)}_{1,2}}}\bigr)\circ\Phi^{(\nu)}_{1,2}\cong\id_{\pi_1^{(\nu)\,*}\cG}\,,
\qqq}
\hspace{-0.1cm}where $\,(\Phi^{(\nu)}_{k,k+1},\om^{(\nu)}_{k,k+1})\equiv\pi^{(\vep_\nu)\,*}_{k,k+1}(\Phi^{\vep^{(\nu)}_{k,k+1}},\vep^{(\nu)}_{k,k+1}\,\om)$,\ and where $\,\pi_1^{(\vep_\nu)}=\iota_1\circ\pi^{(\vep_\nu)}_{1,2}\,$ if $\,\vep^{(\nu)}_{1,2}=+1$,\ and $\,=\iota_2\circ\pi^{(\vep_\nu)}_{1,2}\,$ if $\,\vep^{(\nu)}_{1,2}=-1$.\ The superfield theory is determined by the Dirac--Feynman amplitude $\,\cA_{\rm DF}[\xi]=\exp(\sfi\,S_\si[\xi])\,$ on $\,\cF_\si\,$ in which the action `functional' splits $\,S_\si[\xi]=S_{\rm metr}[\xi]+S_{\rm WZ}[\xi]\,$ into the `metric' term $\,S_{\rm metr}[\xi]=\int_\Si\,\Vol(\Si,\xi^*\txg)\,$ and the `topological' WZ term given by the de\-co\-ra\-ted-surface holonomy $\,\exp(\sfi\,S_{\rm WZ}[\xi])=\Hol_{(\cG,\Phi,(\varphi_\cdot))}(\xi|\G)\,$ of \cite{Runkel:2008gr}.

The higher-supergeometric elements $\,\cG,\Phi\,$ and $\,\varphi_{\vep_\nu}\,$ of $\,\Bgt\,$ are distinguished 0-,\ 1- and 2-cells,\ respectively,\ of $\,\bgrb_\nabla(M)$,\ {\it cp.}\ \cite{Stevenson:2000wj,Waldorf:2007mm}.\ The fundamental property of and the main rationale for physical interest in these objects,\ discussed at length in,\ {\it i.a.},\ \cite{Gawedzki:1987ak,Gawedzki:2002se,Suszek:2011hg},\ is that they canonically determine -- {\it via} cohomological transgression,\ originally proposed in \cite{Gawedzki:1987ak} -- a prequantisation of the above superfield theory and encode a lot of non-trivial information on its (nonperturbative) structure,\ {\it cp.}\ \cite{Suszek:2020rev} for a recent review.

\section{The Trinity:\ Simpliciality,\ Symmetry and Semisimplicity}\label{sec:trinity}

The construction of the poly-phase super-$\si$-model from the previous section features submanifolds of Segal's nerve of the $\Si$-fibred pair groupoid $\,{\rm Pair}_\Si(\widehat\Si)\,$ of the extended worldsheet.\ The nerve is a canonical example of a \emph{simplicial manifold},\ and the identities of order 2 satisfied by the structure maps $\,p^{\vep_\nu}_{k,k+1}\,$ and $\,p_A\,$ are readily seen to \emph{follow} from the elementary simplicial identities obeyed by the face maps of $\,{\rm Pair}_\Si(\widehat\Si)$.\ The order-2 relations between the structure maps $\,\pi^{\vep_\nu}_{k,k+1}\,$ and $\,\iota_A\,$ of the background $\,\Bgt\,$ are a target-space realisation of their worldsheet counterparts.\ They give rise to subsets in another simplicial supermanifold,\ to wit,\ the nerve of the pair groupoid $\,{\rm Pair}(M_0)\,$ of the bulk target space ({\it cp.},\ $\,D^\x\,$ and $\,I_{(\nu)}$).\ While there is no {\it a priori} reason to expect that components of the stratified target $\,M\,$ form a simplicial supermanifold,\ there are important circumstances in which they do:\ This happens,\ {\it e.g.},\ in $\si$-models with topological defects with \emph{induction},\ studied in \cite{Runkel:2008gr} in the context of orbifolding,\ in which defect junctions of valence $\,\nu>3\,$ can be obtained from binary trees of defect junctions of valence 3 in a limiting procedure in which the lengths of all \emph{internal} edges are sent to 0,\ at no cost in the value of the action functional (owing to the topological nature of the defects).\ As a result,\ fusion 2-isomorphisms for the junctions of higher valence are induced,\ through (vertical) composition,\ from the elementary ones for trivalent junctions which define the binary-tree resolutions,\ {\it cp.}\ \cite{Runkel:2008gr} and our heuresis in the previous section.\ Such an intrinsically \emph{semi}-simplicial structure is promoted to a fully fledged simplicial one through adjunction of the identity defect,\ encountered previously,\ whose gerbe-theoretic data $\,\id_\cG\,$ are provided by the bulk gerbe itself \cite{Waldorf:2007mm}.\ This defect can be drawn anywhere in the worldsheet (in particular,\ it can be attached to any defect junction,\ whereby the valence of the junction is increased by 1).\ Its existence suggests the incorporation of the degeneracy maps of a simplicial target.

Our hitherto considerations lead us to the definition of a \emph{simplicial superstring background} with the target given by (a submanifold in) a stratified simplicial supermanifold $\,(M_\bullet,d^{(\bullet)}_\cdot,s^{(\bullet)}_\cdot)\,$ with face maps $\,d^{(n+1)}_i:M_{n+1}\too M_n\,$ and degeneracy maps $\,s^{(n)}_i:M_n\too M_{n+1}\,$ defined for $\,i\in\ovl{0,n+1}\,$ and for all $\,n\in\bN$,\ and subject to the standard simplicial identities.\ The $\,d^{(n+1)}_i\,$ reproduce the previously considered structure maps \emph{uniquely} as $\,(\iota_1,\iota_2)=(d^{(1)}_1,d^{(1)}_0)\,$ and $\,(\pi^{(3)}_{1,2},\pi^{(3)}_{2,3},\pi^{(3)}_{1,3})=(d^{(2)}_2,d^{(2)}_0,d^{(2)}_1)$,\ and -- for $\,v>3\,$ -- also $\,\pi^{(v)}_{1,v}=d^{(2)}_1\circ d^{(3)}_1\circ\cdots\circ d^{(v-1)}_1\,$ and $\,\pi^{(v)}_{k,k+1}=d^{(2)}_2\circ d^{(3)}_2\circ\cdots\circ d^{(v-k)}_2\circ d^{(v-k+1)}_0\circ d^{(v-k+2)}_0\circ d^{(v-1)}_0\,$ for $\,k\in\ovl{1,v-1}$,\ consistently with the identities of order 2 mentioned earlier.\ The $\,s^{(n)}_i\,$ account for the existence of the flat identity (sub-)bi-brane $\,s^{(0)\,*}_0\Phi\equiv\id_\cG:s^{(0)\,*}_0 d^{(1)\,*}_1\cG=\cG\cong\cG=s^{(0)\,*}_0 d^{(1)\,*}_0\cG\ox\cI_{s^{(0)\,*}_0\om}\,$ and allow to write down fusion 2-isomorphisms for defect junctions with identity defect lines attached. 

As simpliciality seems to be favoured by topological defects,\ which are transmissive to the Virasoro currents of a \emph{conformal} $\si$-model,\ the Segal--Sugawara realisation of the Virasoro algebra within the universal enveloping algebra of a Ka\v c--Moody algebra of a simple Lie algebra,\ known,\ {\it e.g.},\ from the study of the WZW $\si$-model,\ suggests a natural direction of enhancement of our construction:\ Focusing on $\si$-models with a rich configurational symmetry ({\it e.g.},\ those with targets given by homogeneous spaces of Lie supergroups),\ we may combine simpliciality with symmetry to further constrain the geometry of the supertarget and the simplicial gerbe over it.\ The point of departure is the identification of the bicategorial realisation of rigid symmetries of the $\si$-model,\ which is readily achieved for symmetries induced by isometries of the metric bulk target $\,(M_0,\txg)$.\ Thus,\ we consider a Lie supergroup $\,\txG_\si\,$ together with actions $\,M_n\la:\txG_\si\x M_n\too M_n,\ n\in\bN$,\ and so also with the fundamental vector fields $\,\cK^n_X=-(X\ox\id_{\cO_{M_n}})\circ M_n\la^*\,$ over the $\,M_n\equiv(|M_n|,\cO_{M_n})$,\ labelled by elements of the tangent Lie superalgebra $\,\ggt_\si\ni X\,$ of $\,\txG_\si$.\ The structure maps $\,\pi^{\vep_\nu}_{k,k+1}\,$ and $\,\iota_A\,$ are assumed $\txG_\si$-equivariant to ensure a natural alignment of the bulk and defect-quiver variations of the action functional engendered by the $\,\cK^n_X$,\ {\it cp.}\ \cite{Runkel:2008gr,Gawedzki:2012fu}. We then say that $\,\txG_\si\,$ is a Lie supergroup of prequantisable rigid (configurational) symmetries of the super-$\si$-model if (i) $\,\Om\,$ is $\txG_\si$-invariant;\ (ii) the action admits a generalised relative (co)momentum $\,\k_\cdot:\ggt_\si\too\Om^1(M_0)\oplus\Om^0(M_1)\,$ such that $\,\txD_{M_\cdot}\k_X=-\imath_{\cK_X}\Om$,\ where $\,\txD_{M_\cdot}\,$ is the relative de Rham differential \cite[Prop.\,2.8]{Gawedzki:2012fu};\ and (iii) it lifts to the higher-geometric components of $\,\Bgt\,$ as a $\txG_\si$-indexed\footnote{We put our discussion in the so-called $\cS$-point picture for the sake of simplicity.} family of 1-isomorphisms $\,\La_g:M_0\la_g^*\cG\cong\cG\,$ and 2-isomorphisms $\,\la_g:(\iota_2^*\La_g\ox\id_{\cI_\om})\circ M_1\la_g^*\Phi\cong\Phi\circ\iota_1^*\La_g\,$ satisfying the coherence identities {\small $\,M_{\nu-1}\la_g^*\varphi_{\vep_\nu}=(\varphi_{\vep_\nu}\circ\id)\bullet(\id\circ\la^{(\nu)}_{g\,1,2})\bullet(\id\circ\la^{(\nu)}_{g\,2,3}\circ\id)\bullet\cdots\bullet(\id\circ\la^{(\nu)}_{g\,\nu-1,\nu}\circ\id)\bullet(\la^{(\nu)}_{g\,\nu,1}\circ\id)\,$} (in which some \emph{canonical} 2-isomorphisms have been dropped for brevity).\ The existence of such a coherent lift is a necessary and sufficient condition for the invariance of the decorated-surface holonomy under the symmetry transformations $\,\xi\mapsto M_\cdot\la_g\circ\xi$,\ and can be read off from \cite[Thm.\,4.4]{Gawedzki:2012fu}.\ The marriage between simpliciality and symmetry thus defined can be neatly established by declaring the structure maps of $\,(M_\bullet,d^{(\bullet)}_\cdot,s^{(\bullet)}_\cdot)\,$ $\txG_\si$-equivariant and,\ accordingly,\ by propagating a \emph{given} bulk symmetry $\,M_0\la\,$ over $\,M_\bullet\,$ to engender a simplicial $\txG_\si$-space with an action $\,M_\bullet\la\ :\ \txG_\si\x M_\bullet\too M_\bullet$.\ This yields \emph{maximally (super)symmetric defects}.

Drawing further motivation from the WZW $\si$-model,\ with its generating bi-chiral loop-group symmetries and the corresponding maximally symmetric bi-branes of \cite{Fuchs:2007fw,Runkel:2008gr,Runkel:2009sp} which are supported over \emph{orbits} of the bulk-group action,\ we come to the final stage of the rather natural structurisation of the super-$\si$-model.\ It boils down to imposing the requirement of \emph{semisimplicity} upon the simplicial target $\txG_\si$-su\-per\-mani\-fold,\ by which we mean a (disjoint-sum) decomposition of $\,(M_\bullet,d^{(\bullet)}_\cdot,s^{(\bullet)}_\cdot,M_\bullet\la)\,$ into orbits of the simplicial $\txG_\si$-action.\ The demand that these support -- as a token of quantum-mechanical consistency of the construction -- a simplicial gerbe described previously is then anticipated to give rise to cohomological superselection rules for the \emph{admissible} orbits in the decomposition,\ which are the only ones that we choose to keep.\ For $\,M_\bullet\,$ topologically nontrivial,\ this is bound to yield powerful constraints on the ensuing target geometry,\ coming from the standard Dirac-quantisation argument.\ Compactness of the target geometry should then result in a rationalisation of the (super)background.

\section{The maximally supersymmetric simplicial Lie superbackgrounds}\label{sec:WZWvsCS}

A prime example of a super-$\si$-model to which the above principles may be applied \emph{constructively},\ and in which their consequences may be explored,\ is the WZW(-type) $\si$-model,\ with the bulk target given by a (Kostant--)Lie supergroup $\,\txG$.\ The latter is endowed with a canonical bi-invariant Cartan 3-cocycle $\,\chi_{\rm C}^q=q\,\tr_\ggt(\theta_{\rm L}\wedge[\theta_{\rm L}\overset{\wedge}{,}\theta_{\rm L}]),\ \ggt\equiv{\rm sLie}\txG\,$ which geometrises,\ in physically interesting cases ({\it e.g.},\ for $\,\txG\,$ even compact and connected,\ and on the super-Minkowski group) and for a suitable choice of the loop charge $\,q\in\bR$,\ as a supersymmetric gerbe $\,\cG_{\rm C}$.\ The target is to be seen as an orbit of an action $\,\txG_0\la\,$ of a subgroup of the group $\,\txG\x\txG\,$ of left and right translations on $\,\txG$.\ The corresponding maximally supersymmetric defects implement the \emph{right} regular action $\,\wp\,$ of $\,\txG\,$ on itself,\ {\it cp.}\ \cite{Fuchs:2007fw,Runkel:2008gr,Runkel:2009sp},\ and so here the stratified target $\,M_\bullet\,$ embeds in the nerve $\,\txN(\txG\rx_\wp\txG)\,$ of the right action groupoid $\,\txG\rx_\wp\txG$.\ The morphism composition of the latter groupoid represents supergroup multiplication $\,\txm_\txG\ :\ \txG\x\txG\too\txG$,\ which admits a gerbe-theoretic realisation in the form of a (generalised) supersymmetric \emph{multiplicative structure},\ instrumental in the construction of the bi-brane for the said multiplicative defect.\ The structure is a simplicial gerbe over Segal's model of the classifying space $\,B\txG\,$ of $\,\txG\,$ (containing $\,\txN(\txG\rx_\wp\txG)\,$ as a simplicial sub-manifold),\ and as such comprises a distinguished 1-isomorphism $\,\cM:\pr_1^*\cG_{\rm C}\ox\pr_2^*\cG_{\rm C}\cong\txm_\txG^*\cG_{\rm C}\ox\cI_{\varrho_{\rm PW}}\,$ over $\,\txG^{\x 2}$,\ written in terms of a Polyakov--Wiegmann 2-form $\,\varrho_{\rm PW}$,\ and a coherent (quasi-)associator 2-isomorphism $\,\a\,$ over $\,\txG^{\x 3}$,\ {\it cp.}\ \cite{Carey:2004xt,Suszek:2022lpf} for details.\ As the defect of interest maps to $\,\txG^{\x 2}\,$ and has structure maps $\,(\iota_1,\iota_2)=(\pr_1,\wp\equiv\txm_\txG)$,\ we may invoke the existence of $\,\cM\,$ in the Wong--Affleck argument of Sec.\,\ref{sec:bicat} to conclude that $\,\cG_{\rm C}\,$ must trivialise over the second cartesian factor $\,\iota_D:D\emb\txG\,$ in the relevant bi-brane geometry $\,\widetilde\iota{}_D:\txG\x D\emb\txG^{\x 2}$.\ This,\ however,\ implies that the bi-brane arises as a product $\,\cB_{\rm maxym}\equiv(\txG\x D,\pr_1,\wp,\widetilde\iota{}_D^*\varrho_{\rm PW}-\pr_2^*\om_D,\Phi)\,$ of another kind of \emph{fusion}:\ $\,\Phi=(\widetilde\iota{}_D^*\cM\ox\id_{\cI_{-\pr_2^*\om_D}})\circ(\id_{\pr_1^*\cG_{\rm C}}\ox\pr_2^*\cT_D^{-1}\ox\id_{\cI_{-\pr_2^*\om_D}})\,$ between $\,\widetilde\iota{}_D^*\cM\,$ and the \emph{boundary} bi-brane $\,\cB_\p\equiv(D,\iota_D,\ast,\om_D,\cT_D)\,$ introduced before.\ By a similar argument \cite{Suszek:2022lpf},\ the existence of $\,\a\,$ turns the problem of constructing $\,\varphi_{\vep_{\nu+1}}\,$ into a search -- over (the second factor in) a disjoint union of $\txG_\nu\la$-orbits within $\,\txG\x\widetilde D{}_\nu\equiv\txG\x(D^{\x\nu}\cap\txm_{1\_\nu}^{-1}(D))\,$ with $\,\txm_{1\_\nu}\equiv\txm_\txG\circ(\txm_\txG\x\id_\txG)\circ\cdots\circ(\txm_\txG\x\id_{\txG^{\x\nu-2}})\,$ -- for \emph{boundary} fusion 2-isomorphisms $\,\varphi^{(\p)}_{\nu+1}:\ox_{i=1}^\nu\,\pr_i^*\cT_D\cong(\txm_{1\_\nu}^*\cT_D\ox\id)\circ\cM_{1\_\nu-1,\nu}\rstr_{\widetilde D{}_\nu}$,\ defined in terms of the 1-isomorphism {\small $\,\cM_{1\_\nu-1,\nu}=\cM_{12\cdots\nu-1,\nu}\circ\cdots\circ(\cM_{12,3}\ox\id)\circ(\cM_{1,2}\ox\id)\,$} in which $\,\cM_{12\cdots k-1,k}=((\txm_{1\_k-1}\x\id_\txG)\circ(\pr_1,\pr_2,\ldots,\pr_k))^*\cM$.\ (Incidentally,\ this gives a physical meaning to the seemingly meaningless concept of `fusion of branes' discussed in the literature \cite{Carey:2005qv}.)

In the quantisation of the $\si$-model determined by the gerbe,\ states are represented by disc Dirac--Feynman amplitudes \cite[Sec.\,4.1]{Gawedzki:1999bq},\ and so it stands to reason that the boundary bi-brane $\,\cB_\p\,$ provides a geometric realisation of the spectrum of the quantum theory.\ Accordingly,\ we may anticipate that fusion 2-isomorphisms carry information on its Verlinde fusion ring.\ We conclude this note with a review of evidence which corroborates these expectations and a recapitulation of conjectures based firmly thereon in the setting of the bosonic WZW $\si$-model.\ For further details,\ as well as novel bicategorial constructions for the Green--Schwarz super-$\si$-model with the super-Minkowskian target,\ we refer the Reader to the extensive study \cite{Suszek:2022lpf}.\medskip

\noindent {\bf The bosonic WZW defect \& another link to the CS theory.} The bulk WZW $\si$-model for the compact 1-connected Lie group $\,\txG\,$ is defined by the Cartan--Killing metric $\,\txg=-6q\,\tr_\ggt(\theta_{\rm L}\ox\theta_{\rm L})\,$ and the Cartan 3-form $\,\chi_{\rm C}^q\,$ with $\,24\pi q\equiv\sfk\in\bN^\x$,\ co-normalised in a manner which ensures non-anomalous conformality of the (quantised) field theory.\ Integrality of the \emph{level} $\,\sfk\,$ implies the existence of a unique (isoclass of) gerbe geometrising $\,\chi_{\rm C}^q\,$ -- the $\sfk$-th tensor power of the Gaw\c{e}dzki--Hitchin--Meinrenken basic gerbe $\,\cG_{\rm C}\,$ over $\,\txG$.\ The rigid symmetries of the bulk theory make up $\,\txG_\si=\txG\x\txG\,$ and lift to $\,\sfN_\bullet(\txG\rx_\wp\txG)\,$ as $\,\txG_n\la:\txG_\si\x\txG^{\x n+1}\ni((x,y),g,h_i)\longmapsto(x\cdot g\cdot y^{-1},$ $\Ad_y(h_i))\in\txG^{\x n+1}$.\ Cohomological arguments localise $\,\cB_\p\,$ over the disjoint union $\,D=\sqcup_{\la\in\sfP^\sfk_+}\cC_\la\,$ of the conjugacy classes $\,\cC_\la=\Ad_\txG(\ee^{2\pi\sfi\,\la/\sfk})\,$ labelled by weights $\,\la\,$ from the fundamental affine Weyl alcove $\,\sfP^\sfk_+\,$ at level $\,\sfk$,\ with $\,\om_\p\rstr_{\cC_\la}\,$ fixed \emph{uniquely} by the bi-chiral loop-group extension of $\,\txG_1\la$,\ {\it cp.}\ \cite{Gawedzki:2002se,Suszek:2022lpf}. The existence of $\,(\cM,\a)\,$ is unobstructed,\ and so $\,\cB_\p\,$ induces the non-boundary maximally symmetric bi-brane as discussed above.\ Thus,\ the connected components $\,\txG\x\cC_\la\,$ of the defect-line target are in a 1-1 correspondence with chiral sectors of the bulk Hilbert space,\ furnishing \emph{integrable} highest-weight representations $\,\cV_{\la,\sfk}\,$ of the Ka\v c--Moody algebra $\,\widehat\ggt{}_\sfk$.

The localisation of $\,\cB_{\rm maxym}\,$ conforms with the `triple-S' argument of Sec.\,\ref{sec:trinity},\ and the last remark strengthens the expectation that the associated inter-bi-brane carries \emph{geometric} information on the Verlinde fusion ring of the WZW $\si$-model.\ Recall that the ring structure is encoded by the multiplicity spaces in the decomposition,\ into the $\,\cV_{\la_3,\sfk}$,\ of the Hilbert space of the \emph{boundary} WZW $\si$-model on a strip with boundaries carrying the data of $\,\cB_\p$,\ or,\ equivalently,\ by the spaces $\,\xcC(\cV_{\la_1,\sfk}\ox\cV_{\la_2,\sfk},\cV_{\la_3,\sfk})\,$ of rank-3 conformal blocks of the (chiral) bulk WZW theory,\ {\it cp.}\ \cite{Gawedzki:2001rm}.\ Each such space is the Hilbert space of the 3$d$ Chern--Simons (CS) theory on the time cylinder $\,\bR\x\bC\sfP^1\,$ over $\,\bC\sfP^1$,\ coupled to vertical Wilson lines $\,\bR\x\{\si_i\},\ i\in\{1,2,3\}\,$ with holonomies along the respective non-contractible loops,\ encircling simply the punctures $\,\si_i$,\ valued in the $\,\cC_{\la_i}$,\ {\it cp.}\ \cite{Gawedzki:1999bq}.\ We may now look for imprints of these structures in the \emph{boundary} component $\,T_{++-}^\p\,$ of the elementary inter-bi-brane geometry $\,T_{++-}=\txG\x T_{++-}^\p\,$ ({\it e.g.}),\ which the `triple-S' argument predicts to be (a subset in) the disjoint union of $\Ad_\txG$-orbits in $\,T_{\la_1,\la_2}^{\la_3}\equiv(\cC_{\la_1}\x\cC_{\la_2})\cap\txm_\txG^{-1}(\cC_{\la_3})\,$ for any $\,\la_1,\la_2\,$ and $\,\la_3$.\ And,\ remarkably,\ we find them!\ Indeed,\ the necessary condition for the existence of $\,\varphi^{(\p)}_3\,$ on $\,T_{\la_1,\la_2}^{\la_3}\,$ is the vanishing of $\,\D_{++-}\om\,$ (which restricts to $\,T^\p_{++-}$),\ and the latter 2-form turns out to be\ldots the partially symplectically reduced presymplectic form on the state space of the CS theory described above.\ Its partial reduction,\ due to Alekseev and Malkin \cite{Alekseev:1993rj},\ with respect to the `pointed' gauge group associated with a homological decomposition of $\,\bC\sfP_{(3)}\equiv\bC\sfP\setminus\{\si_1,\si_2,\si_3\}\,$ relative to a point $\,\si_*\in\bC\sfP_{(3)}$,\ leaves us with the tangent of the residual gauge group $\,[\si_*,\txG]\equiv\txG\,$ as the characteristic distribution of $\,\D_{++-}\om$,\ which is \emph{just} the symmetry group $\,\txG\,$ of the $\Ad_\txG$-orbits in $\,T_{++-}^\p$,\ whose disjoint union composes the \emph{classical} state space of the CS theory in the parametrisation of \cite{Alekseev:1993rj}.\ In the light of the interpretation of the conformal blocks as intertwiners $\,\Hom_{\widehat\ggt{}_\sfk}(\cV_{\la_1,\sfk}\ox\cV_{\la_2,\sfk},\cV_{\la_3,\sfk})\,$ between current-symmetry sectors of the chiral bulk theory\footnote{More accurately,\ we should speak of quantum intertwiners between $\,U_{q(\sfk)}(\ggt)$-modules,\ {\it cp.}\ \cite{Gawedzki:2001rm}.},\ this result is also in keeping with the identification of the \emph{transgressed} fusion 2-isomorphisms transmissive to rigid bulk symmetries (which the WZW ones are,\ {\it cp.}\ \cite{Gawedzki:2012fu}) as intertwiners of the symmetry representations on the (twisted) state spaces fusing at the defect junction,\ {\it cp.}\ \cite{Suszek:2011hg,Suszek:2012ddg}.

The structure of the simplicial WZW target is ultimately determined by the requirement of existence of the 2-isomorphisms $\,\varphi^{(\p)}_\nu$,\ expected to distinguish a subfamily within the disjoint union of $\Ad_\txG$-orbits in the boundary factors $\,T_{\vep_\nu}^\p\subset T_{\vep_\nu}$.\ Taking into account the above highly nontrivial result (which generalises to arbitrary $\,\nu$),\ we are led to the following conjectures: 
\ben
\item The fusion 2-isomorphisms $\,\varphi^{(\p)}_\nu\,$ exist only over manifolds $\,\x_{i=1}^{\nu-1}\,\cC_{\la_i}\cap\txm_{1\_\nu-1}^{-1}(\cC_{\la_\nu})\,$ with non-vanishing Verlinde numbers $\,\dim\,\xcC(\ox_{i=1}^{\nu-1}\,\cV_{\la_i,\sfk},\cV_{\la_\nu,\sfk})\,$ (for conformal blocks of arbitrary rank). 
\item Given such a manifold,\ the number of those $\Ad_\txG$-orbits in its decomposition which support $\,\varphi^{(\p)}_\nu\,$ is given by the corresponding Verlinde number.
\een
One may also anticipate that the fusion 2-isomorphisms of valence $\,\nu>3\,$ are induced from the elementary ones with $\,\nu=3\,$ due to simpliciality of the WZW background,\ and that the inter-bi-brane fusing matrices defined,\ as described in detail in \cite{Suszek:2022lpf},\ by the associator move of \cite{Runkel:2008gr} relating inequivalent such induction schemes for $\,\varphi^{(\p)}_4$,\ are intimately related to the standard fusing matrices of the bulk WZW $\si$-model.\ The first conjecture was corroborated for the special case of $\,\txG={\rm SU}(2)\,$ in \cite{Runkel:2009sp},\ and the last expectation hinges on the highly nontrivial cohomological evidence from the simple-current sector gathered in \cite{Runkel:2008gr} as well as on simple considerations of the topologicality of the maximally symmetric defect.\ Various geometric,\ algebraic and field-theoretic arguments in favour of the second conjecture were given in \cite{Suszek:2022lpf}.

\end{document}